# Magneto-Ionic Hardware Security Primitives: Embedding Data Protection at the Material Level

Irena Spasojevic[1,*], Federica Celegato[2], Alessandro Magni[2], Paola Tiberto[2], Jordi Sort[1,3*]

[1]Departament de Física, Universitat Autònoma de Barcelona (UAB), 08193 Bellaterra, Spain
[2]Advanced Materials and Life Science Divisions, Istituto Nazionale di Ricerca Metrologica (INRiM), Strada delle Cacce 91, 10135 Turin, Italy
[3]Institució Catalana de Recerca i Estudis Avançats (ICREA), Pg. Lluís Companys 23, 08010 Barcelona, Spain

**Abstract**

The Big Data revolution has heightened the demand for robust, energy-efficient security hardware capable of withstanding increasingly sophisticated cyber threats. Conventional encryption schemes, reliant on complex algorithms, are resource-intensive and remain vulnerable. To fortify sensitive information, society needs innovative anti-hacking and anti-counterfeiting technologies that exploit new materials and designs. Here, we present a *magneto-ionic strategy* for *hardware-level security* based on fully selective voltage-controlled $N^{3-}$ ion migration within pre-defined, initially paramagnetic FeCoN dots. This process generates ferromagnetic sublayers of tuneable thickness, resulting in either deterministic (single-domain or vortex) or probabilistic states (with coexisting magnetic configurations and voltage-adjustable probabilities), each exhibiting stochastic orientation and chirality, thereby providing a rich platform for magnetic fingerprinting. This approach enables self-protected primitives, including true random number generators, physical unclonable functions, and in-memory probabilistic inference. The resulting reconfigurable architecture combines tamper resistance, low energy consumption, and scalability, marking a significant leap toward next-generation hardware security rooted in emergent magnetic phenomena.

[*]Email: Irena.Spasojevic@uab.cat, Jordi.Sort@uab.cat

## Introduction

Digital technologies and the "Internet of Things" are rapidly reshaping society, with data creation set to skyrocket to 221 zB by 2026. Managing 'Big Data' presents major challenges in processing speed, volume, energy efficiency, and security. With over 422 million records breached in 2024[1], protecting data from unauthorized access is critical. Conventional security relies on software-based methods (passwords, two-factor authentication, antivirus, encryption, etc.), which generate excess data and increase power consumption. Physical anti-counterfeiting features like holograms, security threads, or watermarks are often easy to clone. More sophisticated physical protection measures, such as electrically erasable read-only memories or battery-backed static random-access memories suffer from high energy consumption[2].

A boost towards enhanced data security can be achieved through stochastic or probabilistic hardware-based operations, which outperform software-generated passwords and algorithmic pseudo-random numbers by being unclonable and adversary-resistant. True Random Number Generators (TRNGs) and Physical Unclonable Functions (PUFs)[3-7] are gaining attraction for their cost-effectiveness and robust protection. TRNGs provide high-quality random bitstreams for cryptography[8], while PUFs leverage inherent randomness for rapid authentication *via* challenge-response pairs (CRPs) and secure off-device key storage[8-11]. Beyond security, randomness is pivotal in emerging computing paradigms[12]. Stochastic computing encodes data as TRN-based bitstreams, whereas probabilistic computing uses probabilistic bits (*p*-bits) to process probability distributions within algorithms. Additionally, randomness is vital in neuromorphic systems, potentially mitigating the von Neumann bottleneck[13].

Magnetic materials offer inherent randomness through effects like magnetoresistance variations (spin dice)[14], telegraphic noise in superparamagnetic tunnel junctions[15,16], or thermally influenced skyrmions and domain wall motion[17-19]. However, challenges remain, including Joule heating, limited reconfigurability, data occupancy, and poor probabilistic control. A promising solution is the use of energy-efficient electric fields, rather than currents, to control magnetism. Magneto-ionic materials enable precise modulation of properties such as magnetization, anisotropy, exchange bias, or spin configurations (*e.g.*, transitions between vortex and single-domain (SD) states[20]) *via* voltage-driven ion motion[21-27]. Magneto-ionics also provides non-volatility (eliminating continuous power needs), semiconductor compatibility, rapid readout, and high endurance.

Despite the pressing need for nanoscale control of magnetic properties, most magneto-ionic research so far has focused on continuous films, leaving patterned systems largely unexplored[20,28,29]. Moreover, prior studies have largely concentrated on deterministic effects, overlooking stochastic/probabilistic phenomena near critical thicknesses, such as voltage-driven transitions between SD and vortex states[20]. Furthermore, existing experimental designs report on effects that occur simultaneously in all the patterned entities, with no demonstrations of selective magneto-ionic actuation in dot arrays.

We propose that, by selectively applying voltage to nanopatterned magneto-ionic elements, stochasticity arising from ion motion fluctuations–coupled with their intrinsic multi-state[30,31] or analog memory capabilities[20]–can be uniquely exploited to realize reconfigurable secret keys, crucial in the context of security breaches. Unlike conventional stochastic systems, where randomness is purely passive, magneto-ionic *p*-bit-based devices could enable externally tuneable probabilistic behaviour, providing a pathway towards dynamically adaptive security architectures for next-generation data protection. Combining magneto-ionic stochastic-probabilistic behaviour with selective voltage actuation could enable modulation of complex magnetic configurations, offering enhanced data security at the material level.

Herein, we introduce and experimentally validate the operational principle of *magneto-ionic hardware security primitives*, leveraging the previously unexplored inherent stochasticity and voltage-tuneable probabilistic behaviour of ion motion in FeCoN patterned dots. Using customized electrical circuits, we apply selective voltage protocols to initially paramagnetic FeCoN units, triggering ion motion that probabilistically drives the system into a ferromagnetic SD or vortex state with random orientation or chirality after AC demagnetization. By controlling magnetic sublayer thickness *via* $N^{3-}$ ion migration, we create reconfigurable device fingerprints for next-generation self-protected systems, including TRNGs, PUFs, and in-memory probabilistic inference. This security-by-materials-design approach surpasses vulnerable password systems, offering enhanced tamper resistance, reconfigurability, and improved energy efficiency.

**Selective magneto-ionic actuation within FeCoN functional units**

**Fig. 1a** illustrates the experimental setup, in which FeCoN dots, 20 nm-thick and 2 μm in diameter, are patterned onto a platinum-coated Si substrate *via* electron beam lithography and sputtering (Methods) and placed in a custom-made electrochemical cell. Applying a gate voltage $V_G$ between the bottom Pt layer and the counter electrode in propylene carbonate (PC) solvent generates a strong electric field, driving $N^{3-}$ ion migration in and out of FeCoN dots depending on the voltage polarity. In agreement with our previous reports[20], negative voltage induces a planar ion migration front, expelling $N^{3-}$ ions from initially paramagnetic FeCoN dots and forming a ferromagnetic FeCo sublayer at the dots' bottom (**Fig. 1a**), with tuneable thickness controlled by gating time. This is confirmed *via* electron energy loss spectroscopy (EELS) mapping of a dot's cross-sectional lamella (**Fig 1b,c**), where iron accumulates and nitrogen depletes at its bottom. Kerr imaging at different magnetic fields (**Fig. 1d-g**) reveals that all the FeCo(N) dots become magnetic after gating at – 25 V, reversing their magnetization *via* vortex formation. However, this experimental configuration does not allow voltage application to specific sections of magneto-ionic bit array–an essential capability for complex data storage architectures, multifunctional roles within a single device, and, as later demonstrated, enhanced data security.

**Fig. 2a** illustrates a proposed device scheme for selective voltage control of magneto-ionic units. A 10 × 10 array of FeCoN dots was prepared on an insulating $SiO_2$-coated Si substrate and electrical contacts were patterned using optical and electron beam lithography, followed by gold sputtering and lift-off (Methods). To ensure electrical contact with each magneto-ionic unit, a Ti(10 nm)/Pt(20 nm) bilayer was deposited prior to FeCoN(20 nm), resulting in a total dot stack thickness of 50 nm (**Supplementary Fig. 1**). Two circuits selectively interconnected sections of the FeCoN arrays, as schematically depicted in **Fig 2a**, with scanning electron microscopy (SEM) images in **Fig. 2b,c**. Each contact was linked to the external circuitry and bridged to one of two gold contact pads (1 and 2). Applying negative voltage between a selected contact pad (working electrode) and the metallic counter electrode in a PC electrolyte triggered $N^{3-}$ ion migration exclusively in the electrically connected FeCoN dots. Representative samples and actuated circuits are detailed in Methods. Kerr imaging in **Fig. 2d** confirms that all FeCoN units were initially paramagnetic (OFF state). In Sample 1, applying $V_G(t_1) = -10$ V for $t_1 = 60$ min to contact pad 1 activated the bottom-left circuit ($ON_1$, circuit A), while the rest remained paramagnetic (**Fig. 2e**). Alternatively, applying $V_G = -10$ V for the same duration to pad 2 selectively activated the top-right circuit ($ON_2$, circuit B, **Fig. 2f**). The SEM image in **Fig. 2g** highlights two neighbouring dots (marked by a yellow rectangle in **Fig. 2c**), indicating a contacted (voltage-treated) and an adjacent non-contacted (as-grown) dots. The high-angle annular dark-field scanning transmission electron microscopy (HAADF-STEM) image of a cross-sectional lamella containing both dots (**Fig. 2h**) and the close-up of one dot's interior (**Fig. 2i**), along with cross-sectional nitrogen EELS mapping (**Fig. 2j,k**) of the region marked with a square in **Fig. 2i**, clearly reveal distinct behaviour between the two dots. The non-contacted dot retains a uniform nitrogen distribution, whereas, in the voltage-gated dot, nitrogen migrates to the PC electrolyte leaving residual nitrogen at the top, evidencing a planar ion migration front–a fingerprint of nitrogen magneto-ionic systems[20,32]. This planar migration front enables precise thickness control over the magneto-ionically induced magnetic layer, crucial for determining the system's ground state (vortex *vs*. SD). Such control is not feasible in oxygen magneto-ionic systems (*e.g.*, $CoO_x$, $FeO_x$)[25,29], where $O^{2-}$ migration forms irregular metallic Fe or Co clusters within paramagnetic oxide matrix, thus highlighting nitrogen magneto-ionics' unique advantages.

**Unveiling probabilistic and stochastic behaviour at the magneto-ionic level**

An intriguing aspect of the magneto-ionic dot array is the interplay between stochastic and probabilistic behaviour, evident when the sample undergoes multiple degaussing cycles using an AC magnetic field (Methods). **Fig. 3a,b** show Kerr images of transverse and longitudinal magnetization components in voltage-treated circuit B (**Fig. 2f**), after transverse degaussing. The transverse and longitudinal signals correspond to the in-plane magnetization components along horizontal and vertical directions, respectively. Remarkably, after degaussing, some dots adopt a SD state with orientation of 0° (state 1) or 180° (state 2) relative to the degauss field, while others form vortices with either clockwise (CW, state 3) or counterclockwise (CCW, state 4) chirality (**Fig. 3c**). Hysteresis loop measurements on individual dots (**Fig. 3d,e**) confirm these states: SD dots exhibit square-like loops, while vortex dots display constricted loops, characteristic of magnetization reversal *via* vortex formation[33]. Interestingly, consecutive degaussing cycles (Degauss 1–4) cause dynamic evolution of dots' states, leading to unexpected SD↔vortex transitions, or reversals in vortex chirality/SD orientation within individual dots (**Fig. 3f-j**).

Assigning numerical values (1–4) to the magnetic states of the dots (**Fig. 3c**) and converting images into sequences of numbers based on their (*x, y*) positions enables analysis of randomness and state correlations after repeated degaussing *via* intra-fractional Hamming distance ($FHD_{intra}$). This is computed by calculating the number of differing elements between each unique pair of degaussed sequences, normalized by sequence length (*i.e.*, the number of active dots) and the total number of unique pairs (Equations 1–4, Methods). For a system with four states, the expected FHD$_{intra}$ between random sequences is 0.75. In our four-state system, FHD$_{intra}$ is 0.505 (**Supplementary Fig. 2**), indicating a probabilistic, rather than entirely stochastic, behaviour. This is also reflected in the asymmetric outcome probabilities observed during enrolment (*i.e.*, 100 degauss/imaging cycles): following long-term negative voltage treatment of circuit B ($t_1$ = 60 min), the SD state appears in 8.7% of cases, while vortex occurs 91.3% of the time.

Considering only SD orientation and vortex chirality–grouping the four possible states into two directionality-based subclasses–yields an average FHD$_{intra}$ of 0.474 (**Fig. 3k**), approaching the ideal value of 0.5 for a fully random binary system, and confirming absence of correlations between degaussed states. Therefore, voltage actuation influences SD *vs.* vortex state probability while maintaining full randomness in chirality and orientation, thus unveiling a balance between control and stochasticity.

Remarkably, examining the states of each individual dot after successive degaussing, reveals two types of magneto-ionic bits: (i) deterministic bits (*d*-bits), which always settle into a single state (SD or vortex), and (ii) probabilistic bits (*p*-bits), in which SD and vortex states coexist with a certain probability. These unique features–probabilistic SD/vortex states with fully stochastic orientation/chirality–enable a new class of hardware security primitives, namely magneto-ionic TRNGs and PUFs, unlocking previously unexplored data-security frontiers.

**True random number generator (TRNG) and magneto-ionic lock**

The randomness in the imprinted orientation/chirality of degaussed states can be harnessed for the generation of TRNs directly by the device, eliminating human intervention and vulnerabilities of software-created passwords. To quantify unpredictability and information content of binary *N*-bit outputs, where *N* represents the number of TRN characters, we calculated the average Shannon entropy per bit $\overline{H_i}$ for *N* = 24 magneto-ionically actuated dots (Equations 5–6, Methods). Given the excellent uncorrelation in orientation/chirality (FHD$_{intra}$ ≈ 0.5), the probabilities $p_i$ of the two sub-classes–*i.e.*, right (CW) or left (CCW) orientation (chirality) after degaussing (**Supplementary Table 1**)–were considered, yielding $\overline{H_i}$ = 0.97, approaching the theoretical maximum of 1, thus indicating near-perfect randomness. The total number of possible *N*-bit TRN sequences (*S*) is determined by total Shannon entropy ($H_{total}$) and the number of active dots *N*, following $S = 2^{Htotal=f(N)}$. For *N* = 24 this gives 9.84×10$^6$ possible TRN combinations. As shown in **Fig. 2e,f**, the number of active dots can be reconfigured *via* voltage gating across several circuits, allowing the generation of TRN streams with tuneable length. Magneto-ionic activation of complementary circuit A (**Fig. 2e**) activates additional 18 dots with $\overline{H_i}$ = 0.99 (**Supplementary Table 2**), increasing the number of available sequences to $S = 2^{41} = 2.21×10^{12}$. The FHD$_{intra}$ of this 18-bit array is 0.493 (**Supplementary Fig. 3**).

A schematic of TRNG application for generating tailored-strength *hardware* passwords, *i.e. magneto-ionic locks*, is shown in **Fig. 3l,m**, where password length is controlled magneto-ionically. Longer passwords provide stronger security, as brute-force time scales exponentially with bit length. In our proof-of-principle design, voltage can be simultaneously applied to preselected dots in nine arrays in total, with varying selective contact architectures (**Fig. 3l,m**). Considering 100 activated dots with $\overline{H_i}$ = 0.98, the number of possible passwords reaches $3.17 \times 10^{29}$. Even with an attacker guessing 1 billion passwords per seconds, brute-forcing would take 10 trillion years, vastly longer than the universe's age ($\approx 10^{10}$ years), rendering brute-force attacks infeasible. Beyond security, this approach also suits stochastic computing, where short random bitstreams enable efficient real-time decision-making and frequent updates with minimal memory usage. Thus, magneto-ionic actuation, coupled with selective activation of specific regions within dot arrays, addresses both long-bit security requirements and short-bit computational efficiency.

**Probabilistic magneto-ionic device fingerprints: reconfiguration and inference**

While a random magnetic orientation/chirality benefits TRNG, stochastic behaviour hinders PUF-based authentication, as physically identical samples with fully unpredictable responses under a given challenge are indistinguishable. Instead, the co-existence of vortex and SD states in certain dots, with probabilities adjustable *via* voltage actuation time, as demonstrated in this section, offers a viable approach for magneto-ionic PUF implementation.

Actuating and degaussing circuit A (**Fig. 4 a-d**) under the same conditions as circuit B yields a similar overall vortex state prevalence, with SD states occurring at 10.7% probability. This happens because prolonged negative gating increases the thickness of the magneto-ionically induced FeCo layer, stabilizing dots preferentially in a vortex state[20,33,34]. Tracking spin configurations of individual dots after successive degaussing allows creation of an enrolled library of SD and vortex probabilities (**Supplementary Table 3**). In circuit A (Sample 1), eleven dots are deterministic (vortex-only, highlighted in cyan), while seven behave as *p*-bits (in grey), fluctuating between SD and vortex states. Considering dots' individual state probabilities and positions, this data can be leveraged to create unique "device fingerprints" akin to PUFs. PUF-based authentication relies on measuring a device's unique physical properties through a sequence of queries ("challenges") and their corresponding responses, forming CRPs. In our case, the challenge involves requesting magnetic configurations from a five-dot selection after degaussing, and the Kerr imaging gives the response bitstream. A particular feature of our system is that, unlike conventional PUFs with fixed responses, *p*-bits introduce probabilistic responses, enhancing unpredictability. To determine how many CRPs can be formed from our experimental data, we consider the total number of challenges and their responses. If all 18 dots in the circuit were deterministic, the number of CRPs would match the number of challenges (Equation 7), totalling 8568. Adding 7 *p*-bits increases this to 41,174 (4.8× stronger PUF). While *p*-bits boost entropy and CRPs, they also add complexity to verification.

In conventional PUFs (using only *d*-bits), each challenge has a unique response, and the Bit Error Rate (BER)–the ratio of mismatched bits across trials and the total number of bits in the response–is low, mainly linked to experimental noise. In our method, *p*-bit authentication uses majority voting across multiple trials, assigning each bit as "vortex" or "SD" based on its most probable state (**Supplementary Table 3**). During verification, the device response is compared to the enrolled library after majority voting, and discrepancies introduce BER, which inherently arises from *p*-bits behaviour. The overall BER after *T* consecutive trials is calculated using probability theory (Equations 8–10) by averaging error probabilities across the five-dot positions within the CRP. Representative examples featuring varying *p*-bit counts are shown in **Table 1** (Sample 1, Circuit A). BER decreases with increasing *T*, as discrepancies statistically average out upon successive iterations. BER increases with the number of *p*-bits in the CRPs. For CRPs containing 4 *d*-bits, BER is 0.006% after 5 iterations, but rises to 9.6% when all five dots are *p*-bits. However, for Sample 1, BER consistently falls below 4% after 27 iterations. BER depends on both the intrinsic state probabilities of each *p*-bit and their number within CRPs.

Since FeCo thickness increases with voltage actuation time[20], further voltage exposure would drive most dots toward vortex states, eventually making them fully deterministic (and predictable), therefore reducing the number of CRPs and PUF uniqueness. Accordingly, we shortened the actuation time, from $t_1$ = 60 min (Sample 1) to $t_2$ = 30 min (Sample 2) and compared circuits with the same A-type geometry. Hereafter, Sample 1 and Sample 2 denote circuit A at $t_1$ and $t_2$ in corresponding samples. Shorter treatment time generates a new set of states (**Supplementary Table 3**, Sample 2), increasing the average SD state probability from 10.7% to 28.2%, providing larger vortex/SD diversity (**Fig. 4e-h**). This trend is expected, as the shorter voltage treatment results in reduced thickness of FeCo sublayer within the FeCoN dots[20,33]. Sample 2 also has more p-bits (9 in total), increasing the number of CRPs to 60,858. However, BERs in Sample 2 for some selected dot combinations are higher than in Sample 1, due to the larger fraction of dots with SD probabilities in the 25–75% range, which tend to increase BER more than dots with SD probabilities <25% or >75% (*i.e.*, more deterministic-like).

Although some five-dot combinations yield appreciable BER, we will now show that certain CRPs still allow sample authentication through "probabilistic inference", leveraging the unique probabilistic behaviour of *p*-bits to enhance security. To identify these, we calculate the probabilistic fractional inter-Hamming distance (pFHD$_{inter}$, Equations 11–12) and select dot combinations with pFHD$_{inter}$ ≈ 0.5, indicating strong uniqueness. Examples are listed in **Table 1**. Typically, these combinations comprise *p*-bits in both samples or a vortex in one sample and a *p*-bit in the other, avoiding simultaneous vortices in both samples, which would hinder sample differentiation. We examine the dot combination comprising five *p*-bits in identical positions in both samples–a configuration that maximizes initial uncertainty and BER. In this case, with just a single iteration, differentiating among the two samples is highly challenging, and incorrect guesses are likely. Despite this, BER progressively decreases with successive iterations (**Fig. 4i**), highlighting the suitability of five *p*-bits as robust device fingerprints. For probabilistic inference we selected one of the two samples (*e.g.*, Sample 1), acquired Kerr images of selected five *p*-bits after each degauss (**Fig. 4j**) and compared their states with the majority voted states in the enrolled libraries of both samples. A match with Sample 1 was confirmed if at least three *p*-bits aligned with their enrolled states. Some tests quickly aligned with Sample 1, while in other cases, a single degauss/imaging iteration led to misclassification (40% correct inference, **Supplementary Table 4**). However, even in those cases, cumulatively averaging the authentication probability over 22–27 iterations increased inference reliability to ~90% (**Fig. 4k**). Since each degauss/imaging procedure takes ~3.5 s, authentication in these intricate cases can be completed in ~1.5 min, being considerably faster otherwise.

The proposed TRNG/PUF/probabilistic inference approach offers several advantages. One is its robustness against external magnetic fields–since the probabilities are embedded in the material, effects of magnetic field are deleted by the subsequent degauss, ensuring vortex/SD probabilities remain unchanged and can only be reconfigured by voltage. Additionally, probabilistic inference is wireless, requiring only AC magnetic fields and Kerr imaging. Electrical connections are needed only to set or reconfigure state probabilities, not for demagnetization or reading. Utilizing *p*-bits instead of *d*-bits transforms a potential limitation into an advantage, enabling a large number of CRPs while keeping dot count low, thus minimizing volume occupancy. Importantly, *p*-bits prevent univocal identification from a single or few Kerr images, preventing full data reconstruction from a single interception. This enhances security by concealing data, making it significantly harder for attackers to replicate or manipulate authentication patterns, thereby adding an additional layer of protection against hacking and counterfeiting. The system also provides tamper resistance by detecting cyberattacks involving unauthorized voltage application. Since only some dots are initially activated (ON) while others remain OFF, an intruder applying voltage incorrectly would inadvertently switch non-active dots, exposing the intrusion and triggering an alert for reconfiguration. Finally, *p*-bits security measures may help counteract brute-force and quantum hacking attacks.

**Conclusions**

We demonstrate the selective actuation of FeCoN units within designed circuits, enabling targeted magneto-ionic transformations that convert initially paramagnetic FeCoN into vortex or SD states *via* voltage application. We unravel the probabilistic and stochastic nature of magneto-ionics, demonstrating that while some dots consistently generate deterministic states (*d*-bits) under voltage gating, others behave as *p*-bits allowing vortex and SD states to coexist with a defined probability. Voltage gating provides a means to tune the overall vortex/SD probability and thus *d*-bit–to–*p*-bit ratio, allowing strategic manipulation of the device fingerprints while preserving the crucial balance between control and stochasticity, essential for data security applications. This allowed us to develop "self-protected" magneto-ionic data security primitives. We show that the random orientation/chirality of generated states can be exploited to create TRNs and hardware cryptographic keys, whereas the vortex/SD probabilities at each dot position within the array can serve as unique device identifiers. We demonstrate a probabilistic inference approach that permits sample identification though an enrolment/verification process *via* probabilistic PUF scheme, analogous to training/inference in neuromorphic in-memory computing systems. These advancements are poised to revolutionize secure hardware authentication, cryptography, and anti-counterfeiting, enabling reconfigurable and tamper-resistant security systems. Moreover, they pave the way for multifunctional devices capable of parallel operations *via* selective actuation protocols, supporting unconventional computing paradigms and even quantum randomness generation–all powered by the unique capabilities of selective magneto-ionic actuation.

**Methods**

**Sample preparation**

Sample preparation comprised three successive lithography steps, which involved patterning of FeCoN dots, lithography of microscopic selective contacts and finally preparation of macroscopic contacts. 2 µm-diameter Ti (10 nm)/Pt (20 nm)/FeCoN (20 nm) dots, whereby the number in parenthesis denotes thickness of a given layer, were prepared by electron beam lithography (EBL) and subsequent magnetron sputtering. A single layer PMMA 4A photoresist was deposited by spinning for 60 s at 6000 rpm and baked for 5 min at 438 K on top of Si/SiO$_2$ (1.5 µm) substrate. EBL was performed by using an environmental scanning electron microscope ESEM FEG Quanta 3D™ equipped with NanoPattern generator system for nanolithography. Photoresist was developed using 1:3 methyl isobutyl ketone/isopropanol (MIBK/IPA) for 60 s. Bottom Ti/Pt conductive bilayer was initially deposited *via* DC sputtering using an AJA International ATC 2400 sputtering system in Ar atmosphere at a total pressure of $3\times10^{-3}$ Torr, by channelling the corresponding targets to a DC source at 200 W and 100 W power, respectively. A 20 nm thick layer of ternary nitride Fe$_{0.65}$Co$_{0.35}$N (FeCoN) was consecutively deposited through reactive magnetron co-sputtering at room temperature in a mixed Ar/N$_2$ atmosphere and at a total pressure of $3\times10^{-3}$ Torr[32]. The Ar:N$_2$ flow ratio was set at 1:1 to produce nitrogen-rich dots with paramagnetic properties. The Fe target was powered at a constant 50 W using DC, while the Co target was powered at 55 W using an RF source. Finally, the unexposed photoresist was removed *via* a lift-off process in an ultrasonic acetone bath, leaving behind Ti/Pt/FeCoN dots on top of Si/SiO$_2$ substrate. In the case of sample without selective contacts (Fig. 1), FeCoN dots were grown following the same procedure on top of continuous Ti (10 nm)/Pt (20 nm) layer, initially deposited on top of Si substrate. For selective contact architectures proposed here, microscopic contacts with a width of $\approx$ 500 nm used to selectively interconnect magneto-ionic FeCoN units, were prepared by EBL following the same procedure previously described for dots, involving deposition of gold and consecutive lift-off. Macroscopic contacts were prepared *via* optical lithography. First, image reversal photoresist AZ5214E was spin-coated at 3500 rpm for 30 s onto the prefabricated microscopically interconnected Si/SiO$_2$/Ti/Pt/FeCoN sample, followed by a prebake at 358 K for 60 s to ensure proper adhesion and resist activation. Using a mask aligner, a previously designed metallic mask shaped for macroscopic contacts was spatially aligned with the interconnected FeCoN arrays and exposed to UV radiation for few seconds, followed by the reversal bake for 60 s at 358 K. After the reversal bake, the areas exposed to UV light become cross-linked and insoluble in the developer, while the unexposed areas still behave like a normal unexposed positive photoresist. After a flood exposure (without the mask), these areas got dissolved in AZ351 developer, while the cross-linked areas remained, therefore giving a negative image of the mask pattern. Patterning of macroscopic contacts is followed by the sputtering of gold and a final lift-off process in acetone, giving a complete device. Representative samples and actuated circuits commented in the main text are detailed below:

| Sample | Circuit | Gating voltage | Gating time |
|---|---|---|---|
| **Sample 1** | Circuit A | – 10 V | 60 min |
| | Circuit B | – 10 V | 60 min |
| **Sample 2** | Circuit A | – 10 V | 30 min |

During the magneto-ionic actuation to activate selective circuit of dots, a gate voltage is applied between the working electrode (contact pad) and the counter electrode (Pt wire) immersed in PC solvent.

**Magneto-optic Kerr imaging**

Kerr imaging was carried out using a high-resolution microscope (Evico Magnetics) equipped with an oil immersion objective lens. An in-plane magnetic field was applied using a rotatable electromagnet with split pole pieces to enhance field uniformity. This setup allows for the simultaneous acquisition of two perpendicular components of the magnetization: transverse (parallel to the applied field) and longitudinal (perpendicular to the applied field). The sample was demagnetized from a maximum field of 300 Oe, with the field oscillating at a frequency of 23 Hz and linearly decaying over 3 s, followed by Kerr imaging. Images were acquired at 30 ms/frame (~33 fps) with binning 2 or 120 ms/frame (~8 fps) with binning 1. Final images were obtained by averaging 16 frames, yielding effective acquisition times of ~0.5 s (binning 2) and ~2 s (binning 1). In Kerr imaging, the vortex state appears as alternating dark and bright contrast dividing the dot into two distinct regions and indicating the in-plane magnetization orientation (Fig. 3c). By simultaneously obtaining transverse and longitudinal magnetization component, the chirality of the vortex can be determined. In contrast, the SD state exhibits unidirectional magnetization, resulting in a uniform magnetic contrast (black, white, or grey), depending on its orientation.

**Electron microscopy**

Scanning electron microscopy (SEM) imaging was performed using a Zeiss Merlin SEM microscope. High-angle annular dark-field scanning transmission electron microscopy (HAADF-STEM) and electron energy loss spectroscopy (EELS) were performed on a FEI TECNAI G2 F20 HRTEM/STEM microscope with a field emission gun operated at 200 kV. Cross-sectional lamellae of the samples were cut by focused ion beam after the deposition of Pt protective layers and were subsequently placed onto a Cu TEM grid.

**Atomic Force Microscopy (AFM)**

Atomic force microscopy measurements of samples' topography were conducted in tapping mode using the MFP-3D Origin+ Atomic Force Microscope from Asylum Research, Oxford Instruments.

### The intra-fractional Hamming distance (FHD$_{intra}$)

FHD$_{intra}$ after multiple degausses is determined by counting the number of differing elements between each unique pair of degaussed state sequences and then normalizing by the sequence length (*i.e.*, the number of active dots, $N$) and the total number of unique pairs.

The Hamming distance[35] between two sequences $S_i$ and $S_j$ (each of length $N$) can be calculated as:

$$HD(S_i, S_j) = \sum_{k=1}^{N} \left(1 - \delta_{S_k^i, S_k^j}\right) \quad (1)$$

where $\delta_{S_k^i, S_k^j}$ is the Kronecker delta function and $S_k^i$ and $S_k^j$ are the $k^{th}$ elements of the sequences $S_i$ and $S_j$, respectively. This counts the number of positions where two sequences differ.

The number of unique pairs of degaussed images is given by the combination formula:

$$Total\ pairs = \binom{M}{2} = \frac{M(M-1)}{2} \quad (2)$$

Where $M$ is the total number of acquired degaussed images.

Thus, the total sum of Hamming distances over all pairs is:

$$HD_{total} = \sum_{i=1}^{M} \sum_{j=i+1}^{M} HD(S_i, S_j) \quad (3)$$

The intra-fractional Hamming distance is given by:

$$FHD_{intra} = \frac{2}{M(M-1)N} \sum_{i=1}^{M} \sum_{j=i+1}^{M} HD(S_i, S_j) \quad (4)$$

### Shannon entropy

To quantify the average unpredictability and the information content of binary $N$-bit outputs–where $N$ represents the number of TRN characters corresponding to the number of magneto-ionically actuated dots–we calculate the Shannon entropy for the displayed array of $N = 24$ dots as[36]:

$$H_{total} = -\sum_{i=1}^{N} H_i = -\sum_{i=1}^{N} p_i log_2 p_i + (1 - p_i) log_2 (1 - p_i) \quad (5)$$

Where $H_i$ is entropy contribution of $i$-th bit. Due to the excellent uncorrelation in the orientation/chirality of the degaussed states (FHD$_{intra}$ ≈ 0.5), here we consider the probabilities $p_i$ of obtaining either a right (CW) or left (CCW) orientation (chirality) after degaussing (**Supplementary Table 1, 2**), respectively. Therefore, the total Shannon entropy of our system is expressed as:

$$H_{total} = -\sum_{i=1}^{N} p_{R,CW_i} log_2(p_{R,CW_i}) + p_{L,CCW_i} log_2(p_{L,CCW_i}) \qquad (6)$$

**Number of challenge-response pairs (CRPs)**

Using probabilistic theory, the number of CRPs can be determined from the following equation:

$$no.\,CRP = \sum_{m=0}^{min(P,k)} \binom{P}{m}\binom{D}{k-m} \cdot 2^m \qquad (7)$$

where $P$ is the total number of p-bits in the actuated circuit, $D$ is the total number of deterministic bits (so that $P + D = N$, being $N$ the total number of actuated dots), $m$ is the number of p-bits within the $k$-digit CRP, and $k - m$ is the number of deterministic bits within the $k$-digit CRP. Note that $\binom{P}{m}\binom{D}{k-m}$ represents the total number of challenges for a given m, while $2^m$ represents the corresponding total number of responses. For $m = 0$ (all bits in the $k$-digit CRP are deterministic, then the number of responses is 1. Also note that $\binom{P}{m}$ is a factorial number that can be calculated as: $\binom{P}{m} = \frac{P!}{m!(P-m)!}$.

**Bit Error Rate (BER)**

To compute the error probability after majority voting, one can calculate, for each dot, the probability that more than half of the trials are wrong. The probability of $k$ errors in $T$ trials follows a binominal distribution[37], given by:

$$p(k) = \binom{T}{k}(p_e)^k(1-p_e)^{T-k} \qquad (8)$$

where $p_e$ is the probability of error in that dot, and $(1 - p_e)$ is the probability of success. The final error probability for that dot after majority voting, $p_e^{majority}$, is the sum of probabilities where the number of errors, $k$, exceeds more than a half of $T$, i.e., $k = \lfloor T/2 \rfloor + 1$. That is:

$$p_e^{majority} = \sum_{k=\lfloor T/2 \rfloor+1}^{T} p(k) \qquad (9)$$

Finally, taking into account the error probabilities of all *N* dots, BER is given by:

$$BER = \frac{\sum_{i=1}^{N} p_e^{majority}(i)}{N} \qquad (10)$$

**Inter-fractional Hamming distance (FHD$_{inter}$) in the context of probabilistic inference**

The uniqueness of given combinations of dots when comparing two samples can be evaluated through the calculation of the FHD$_{inter}$. For this, the states (vortex *vs*. SD) can be compared dot by dot. Since the probabilities of having vortex or SD for each dot in the array are known, the FHD$_{inter}$ can be calculated probabilistically by considering the likelihood of mismatch between corresponding dots in two samples based on their respective state[38]. For each pair of dots at position *k* in Sample 1 and Sample 2, the mismatch probability, $p_{mismatch}$, can be estimated as:

$$p_{mismatch,k} = p_k^{V1} p_k^{SD2} + p_k^{SD1} p_k^{V2} = p_k^{V1} + p_k^{V2} - 2 p_k^{V1} p_k^{V2} \qquad (11)$$

Where $p_k^{V1}$ and $p_k^{SD1}$ are the probabilities of vortex or SD state in Sample 1 at *k*-th position, while $p_k^{V2}$ and $p_k^{SD2}$ correspond to those in Sample 2.

Then, probabilistic fractional inter-Hamming distance (*p*FHD$_{inter}$) is given as:

$$pFHD_{inter} = \frac{1}{N} \sum_{k=1}^{N} p_{mismatch,k} \qquad (12)$$

Where, in our case, *N* is the total number of dots within the CRP.

**Data availability**

The authors declare that the data supporting the findings of this study are available within the paper and its supplementary information files. Source data have been deposited in the Figshare database.

**Acknowledgements**

Financial support by the European Research Council (2021-ERC-Advanced REMINDS Grant Nº 101054687), the *Generalitat de Catalunya* (2021-SGR-00651), and the Spanish Government (PID2020-116844RB-C21) are acknowledged. Views and opinions expressed are however those of the author(s) only and do not necessarily reflect those of the European Union or the European Research Council Executive Agency. Neither the European Union nor the granting authority can be held responsible for them. This work/Part of this work has been carried out at Nanofacility Piemonte, a laboratory supported by the ''Compagnia di San Paolo'' Foundation, and at QR Lab - Micro&Nanolaboratories, INRiM.

**Author Contributions**

J.S. and I.S. conceived the study. J.S. supervised the study. I.S. and F.C. prepared the samples and A.M. performed Kerr imaging measurements with inputs from F.C., P.T., I.S. and J.S. I.S., J.S. and A.M. analysed the results. I.S. and J.S. performed HR-TEM, HAADF-STEM and EELS measurements and analysed the results. I.S. performed AFM measurements. F.C. performed SEM measurements. I.S. wrote the manuscript with input from J.S. All authors contributed to the discussions.

**Competing interests**

A patent related to this work has been filed by inventors I.S., F.C., A.M., P.T. and J.S.

**Corresponding authors**

Correspondence to Irena Spasojevic and Jordi Sort.

Figures

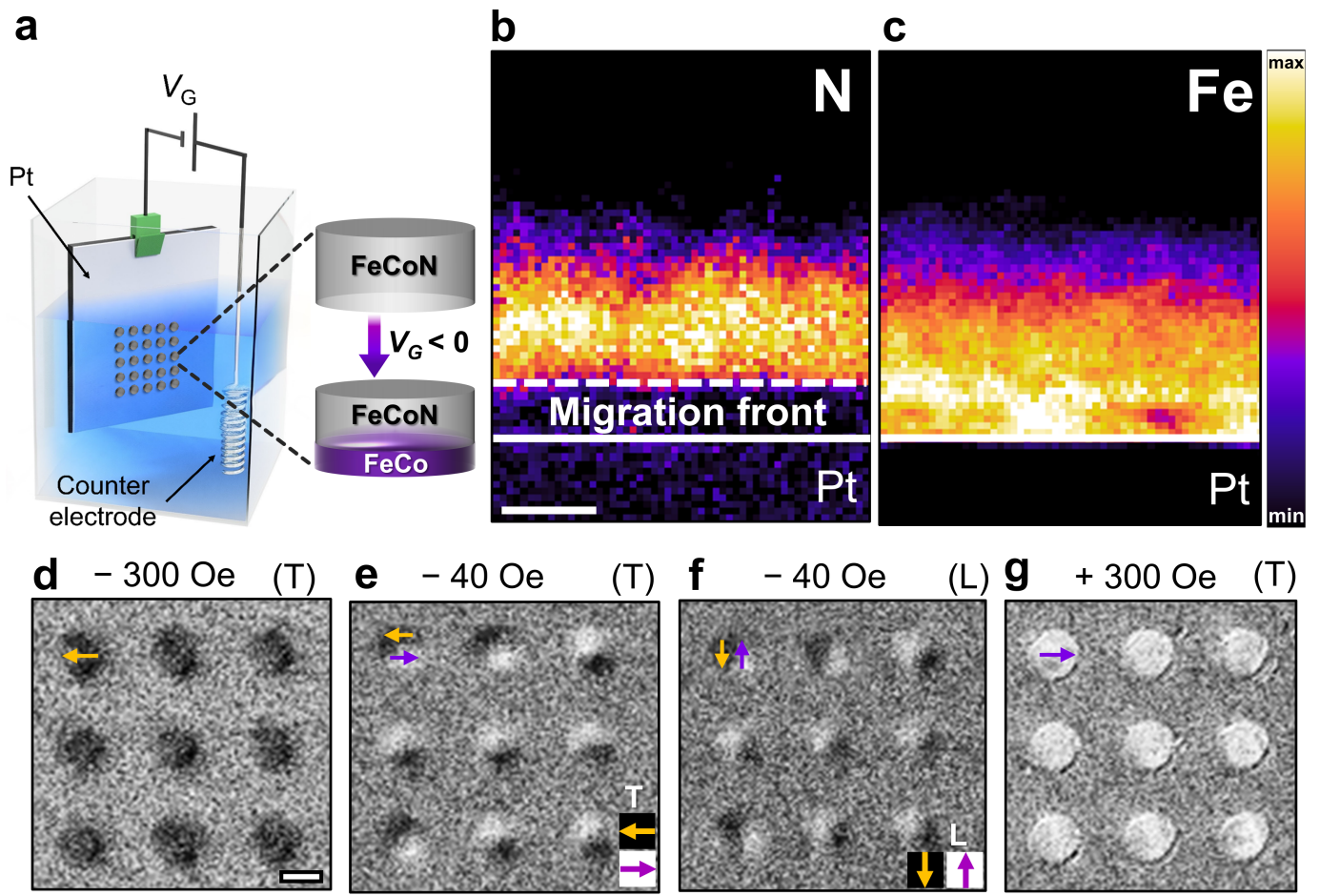

**Figure 1. Magneto-ionic actuation and Kerr imaging of FeCoN dots. a**, Schematic of the custom-built electrochemical cell illustrating voltage-driven $N^{3-}$ ion migration in paramagnetic FeCoN dots at gate voltage $V_G < 0$, leading to the formation of a magnetic FeCo layer at their base. EELS compositional mapping of nitrogen (**b**) and iron (**c**) ions in a dot after negative voltage gating, revealing a planar nitrogen migration front (white dashed line) that divides the dot into two sublayers: nitrogen-rich, paramagnetic FeCoN at the top, and nitrogen-free, ferromagnetic FeCo at the bottom. Scale bar in (**b**): 10 nm. **d–g**, MOKE imaging of a 3 × 3 dot array in transverse (T) geometry at − 300 Oe (*i.e.* negative saturation) (**d**), − 40 Oe (**e**) and (**g**) + 300 Oe (*i.e.*, positive saturation), demonstrating magnetization reversal *via* vortex formation in all the dots. The presence of vortex states with clockwise (CW) or counterclockwise (CCW) chirality is evident from the dark-bright contrast in transverse T (**e**) and longitudinal L (**f**) Kerr images, indicating the curling of magnetic moments around a central core. Scale bar in (**d**): 2 µm.

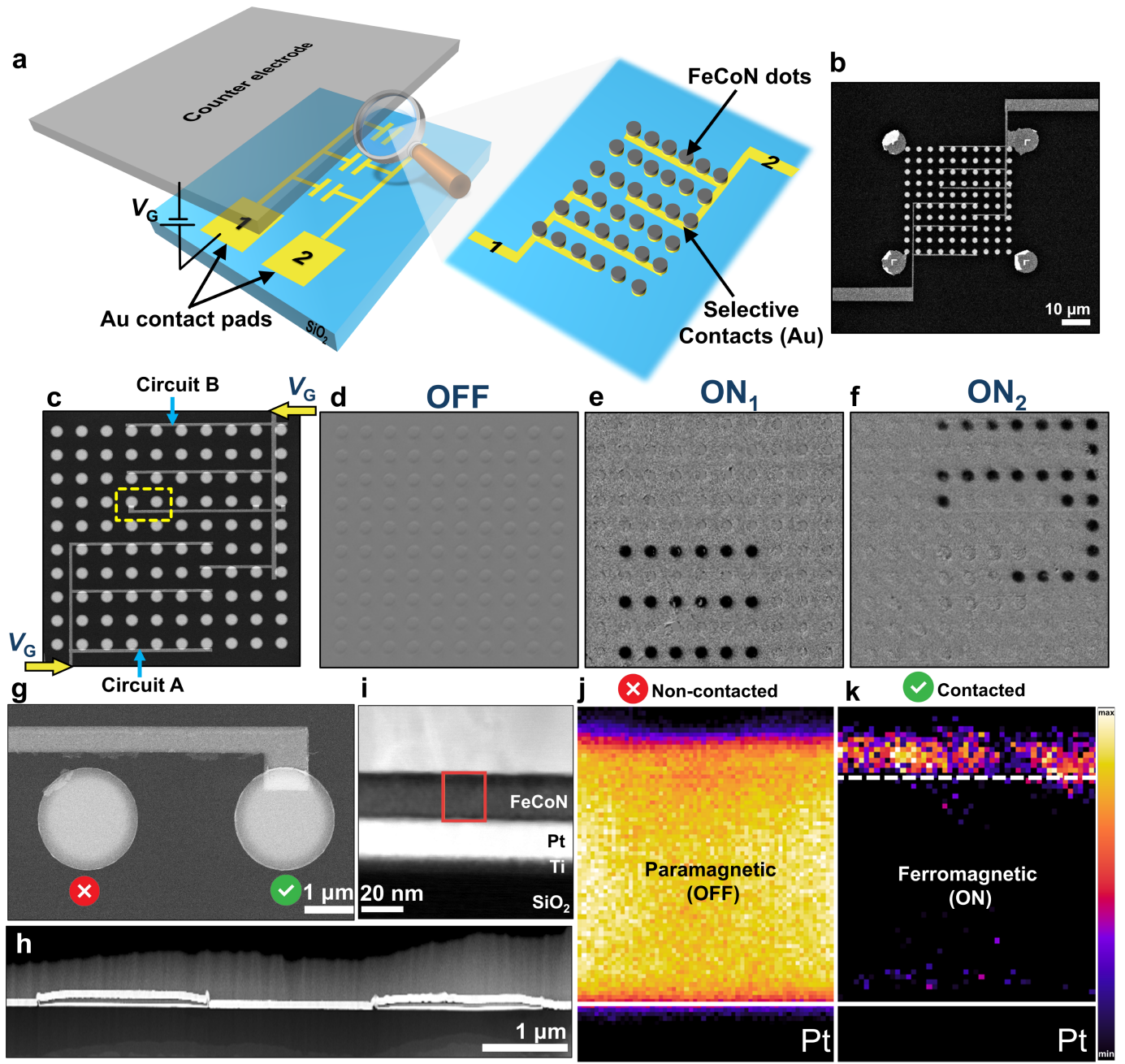

**Figure 2. Operating principle and prototype device for selective magneto-ionic actuation of individual circuits. a**, Schematic representation of a device, whereby application of gate voltage $V_G < 0$ between the counter electrode and either contact pad (labelled 1 and 2) selectively activates dots within circuit A or circuit B *via* precisely defined lithographed contacts. **b**, SEM image of two nano-sized electrical pathways interconnecting sections of a 10 × 10 dot array, along with two larger electrical contacts on each side for integration with macroscopic circuitry. **c**, Details of the patterned selective contacts interconnecting bottom left (circuit A) and top right (circuit B) part of the array. **d**, Kerr imaging of the interconnected FeCoN dot array before voltage application, showing all FeCoN units in initial paramagnetic (OFF) state. **e**, Kerr image obtained under a magnetic field of – 300 Oe (*i.e.*, at saturation) after negative voltage gating of circuit A, causing initially paramagnetic FeCoN dots to become

ferromagnetic (state $ON_1$), while the remaining dots stay paramagnetic. **f**, Application of a negative gate voltage to circuit B, which selectively activates the top-right section of the array (state $ON_2$). In both cases gate voltage $V_G = -10$ V was applied for $t_1 = 60$ min. **g**, SEM image of two dots, marked by a yellow rectangle in Fig. 2**c**, where the contacted dot (green tick) underwent negative voltage treatment, while the non-contacted dot (red cross) remains in as-grown paramagnetic state. **h**, HAADF-STEM image of a cross-sectional lamella containing both dots, with a close-up of one dot's interior in (**i**). **j,k**, EELS nitrogen compositional mapping of non-contacted (*i.e.* as grown, paramagnetic – OFF, **j**) and voltage treated (ferromagnetic – ON, **k**) dot.

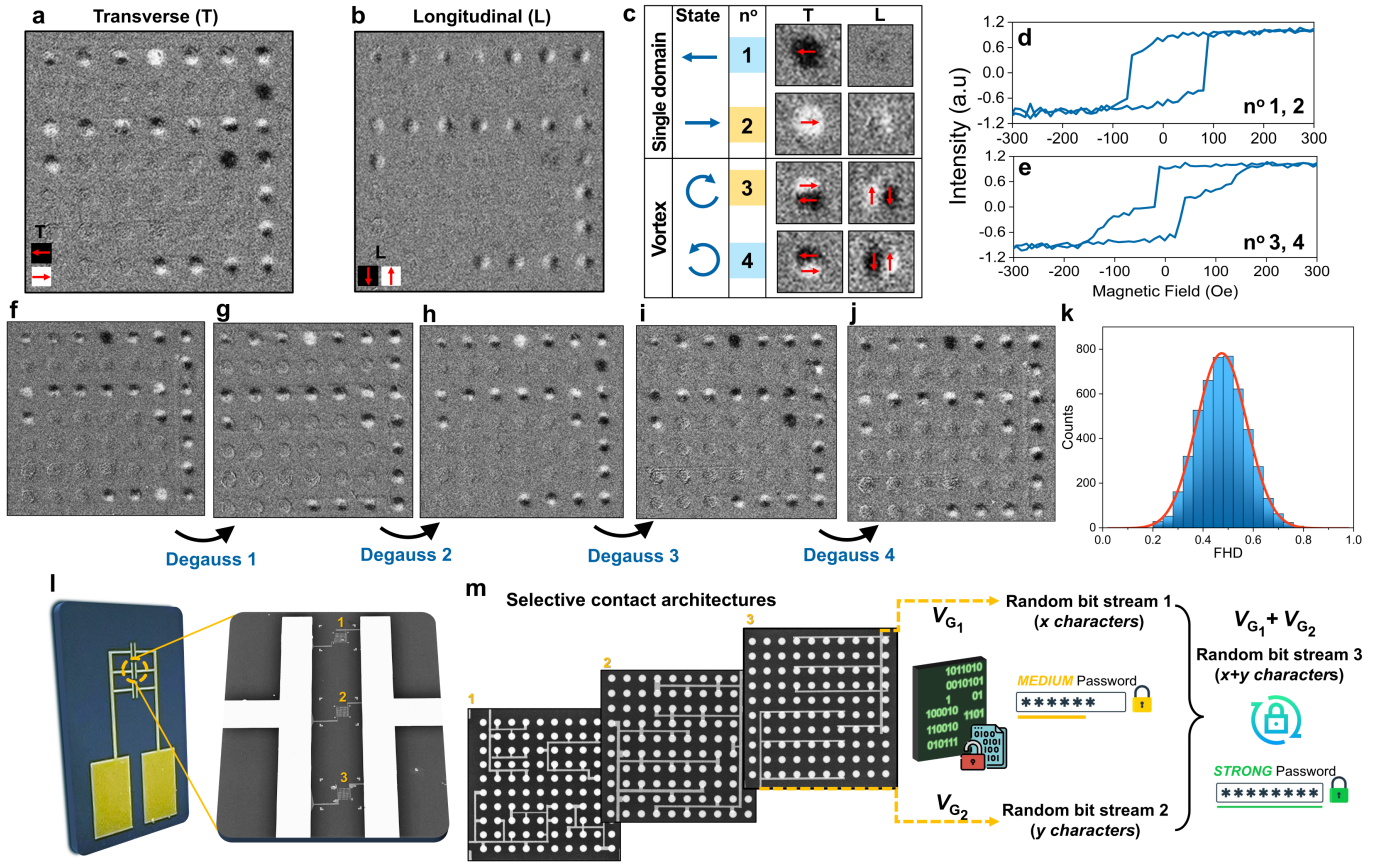

**Figure 3. Stochastic behaviour at the magneto-ionic level – magneto-ionic lock.** Kerr imaging of the transverse T (**a**) and longitudinal L (**b**) magnetization components of dots in voltage-treated circuit B in Sample 1 (shown in **Fig. 2f**) after AC degaussing in the transverse direction. The latter reveals two distinct spin configurations–single-domain (SD) or vortex– with varying directions and chiralities, giving a total of four states depicted in (**c**). Hysteresis loops reconstructed from Kerr imaging contrast at various applied magnetic fields for single-domain (**d**) and vortex (**e**) states. **f-j**, Transverse Kerr images obtained after successive degaussing, revealing probabilistic transitions between SD and vortex states, along with stochastic changes in SD states' directionality and vortex chirality. **k**, Intra fractional Hamming distance (FHD$_{intra}$) between multiple degaussing cycles, considering only SD state orientation and vortex state chirality (*i.e.* two subclasses based on directionality obtained by grouping states 1+4 and 2+3). The red line represents the Gaussian fit of the histogram, yielding an average FHD$_{intra}$ = 0.474, indicating truly random and uncorrelated degaussed states. Distinct selective contact architectures (**l**) and envisioned application of demonstrated true random number generator for generating tailored-strength *hardware* passwords (**m**), *i.e. magneto-ionic locks*, where the number of password characters is controlled magneto-ionically by activating one or more selective contacts.

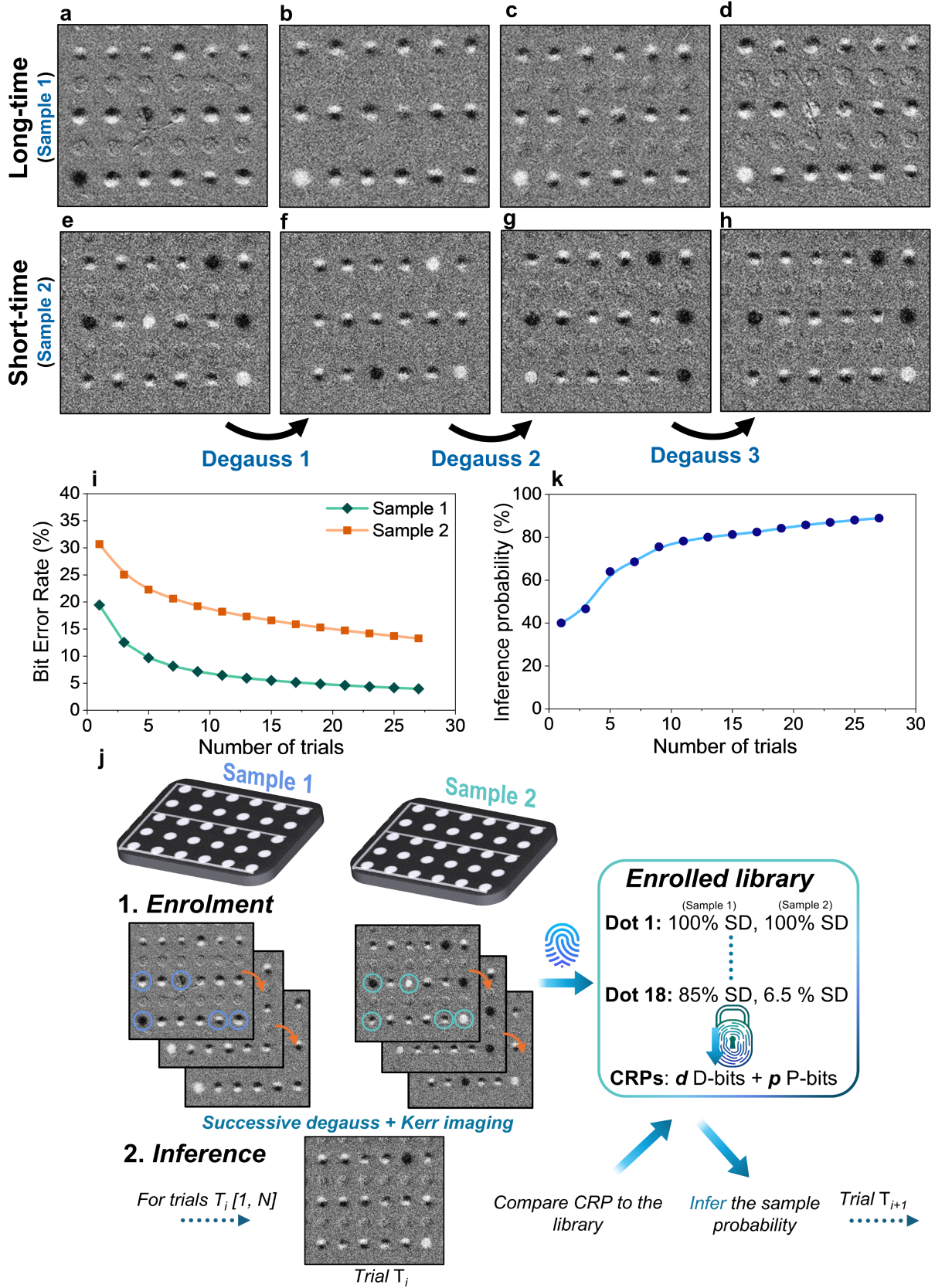

**Figure 4. Probabilistic magneto-ionic device fingerprints: reconfiguration and inference.** Transverse Kerr images obtained after AC degaussing and voltage treatment at $V_G = -10$ V of (**a-d**) circuit A for $t_1$ = 60 min (Sample 1, "long-time") and (**e-h**) circuit of the same geometry, treated for $t_2$ = 30 min (Sample 2, "short-time"). Decreasing the gating time increases the number of *p*-bits (see also related **Table 1** and **Supplementary Table 3**) and brings their SD/vortex probabilities closer together. **i**, Evolution of Bit Error Rate (BER) as a function of number of trials for the combination of five selected *p*-bits for both samples. **j**, Probabilistic inference protocol, where Kerr images of five selected *p*-bits (*i.e.* challenge-response pair, CRP) are acquired after each degaussing of an unknown sample (*e.g.*, Sample 1 in this case) and compared with the "majority-voted" states from the enrolled libraries of both samples to infer whether the sample in question is Sample 1 or Sample 2. Even with the high inherent uncertainty when five *p*-bits are used, cumulatively averaging the authentication probability over 22–27 iterations ensures reliable inference, achieving ~90% inference certainty (**k**).

| Circuit A | Sample 1 | | | | | Sample 2 | | | | | |
|---|---|---|---|---|---|---|---|---|---|---|---|
| Dot positions | No. p-bits | $BER_1$ (%) | $BER_5$ (%) | $BER_{11}$ (%) | $BER_{27}$ (%) | Dot positions | No. p-bits | $BER_1$ (%) | $BER_5$ (%) | $BER_{11}$ (%) | $BER_{27}$ (%) | $pFHD_{inter}$ |
| 5,11,12,14,15 | 1 | 0.640 | 0.006 | 0.000 | 0.000 | 5,11,12,14,15 | 3 | 18.020 | 10.760 | 6.814 | 3.599 | 0.5002 |
| 5,11,13,15,17 | 2 | 4.900 | 0.685 | 0.062 | 0.000 | 5,11,13,15,17 | 4 | 30.020 | 22.660 | 17.270 | 10.870 | 0.4921 |
| 1,5,12,17,18 | 2 | 4.440 | 0.610 | 0.060 | 0.000 | 1,5,12,17,18 | 3 | 14.760 | 7.360 | 4.160 | 1.730 | 0.5999 |
| 7,10,11,12,18 | 3 | 8.400 | 2.193 | 0.601 | 0.035 | 7,10,11,12,18 | 4 | 21.800 | 12.700 | 8.780 | 6.120 | 0.4999 |
| 5,9,11,13,17 | 3 | 13.280 | 7.700 | 5.862 | 3.924 | 5,9,11,13,17 | 4 | 24.160 | 16.200 | 11.860 | 7.400 | 0.4975 |
| 3,7,10,14,18 | 4 | 9.040 | 2.200 | 0.600 | 0.040 | 3,7,10,14,18 | 3 | 13.180 | 8.710 | 7.390 | 5.990 | 0.3065 |
| 7,9,14,15,18 | 4 | 15.180 | 8.980 | 6.390 | 3.960 | 7,9,14,15,18 | 4 | 22.580 | 15.680 | 12.830 | 9.460 | 0.4421 |
| 7,9,13,17,18 | 5 | 19.440 | 9.688 | 6.548 | 3.959 | 7,9,13,17,18 | 5 | 30.680 | 22.270 | 18.210 | 13.270 | 0.5086 |

**Table 1. Some representative five-dot combinations for challenge-response pairs in Sample 1 and Sample 2 (circuit-type A) along with calculated Bit Error Rate (BER) and probabilistic fractional inter-Hamming distance (pFHD$_{inter}$).** For each sample, the position of the dots in the array is identified by a number from 1 to 18. The dots are numbered following the reading direction, from left to right and top to bottom, so that dot number 1 is at the upper-left corner, and dot number 18 is at the bottom-right corner of circuit A in both samples. The second column indicates the number of *p*-bits in each CRP (No. *p*-bits), and the exact positions of the *p*-bits in the circuit are highlighted in bold black within the CRP. Deterministic bits (*d*-bits) that consistently adopt vortex states are highlighted in green, while those that always adopt a SD state are marked in red. The bit error rate (BER), calculated by comparing the device response to the enrolled library after majority voting over a specified number of trials (*T*), is shown for *T* = 1, 5, 11, and 27, with *T* indicated in the subscript. Dot combinations with calculated pFHD$_{inter}$ close to 0.5 indicate strong uniqueness and are well-suited for the sample differentiation *via* probabilistic inference.

# Supplementary Information

**Magneto-Ionic Hardware Security Primitives: Embedding Data Protection at the Material Level**

Irena Spasojevic[1,*], Federica Celegato[2], Alessandro Magni[2], Paola Tiberto[2], Jordi Sort[1,3*]

[1]Departament de Física, Universitat Autònoma de Barcelona (UAB), 08193 Bellaterra, Spain
[2]Advanced Materials and Life Science Divisions, Istituto Nazionale di Ricerca Metrologica (INRiM), Strada delle Cacce 91, 10135 Turin, Italy
[3]Institució Catalana de Recerca i Estudis Avançats (ICREA), Pg. Lluís Companys 23, 08010 Barcelona, Spain

[*]Email: Irena.Spasojevic@uab.cat, Jordi.Sort@uab.cat

# I. Atomic Force Microscopy

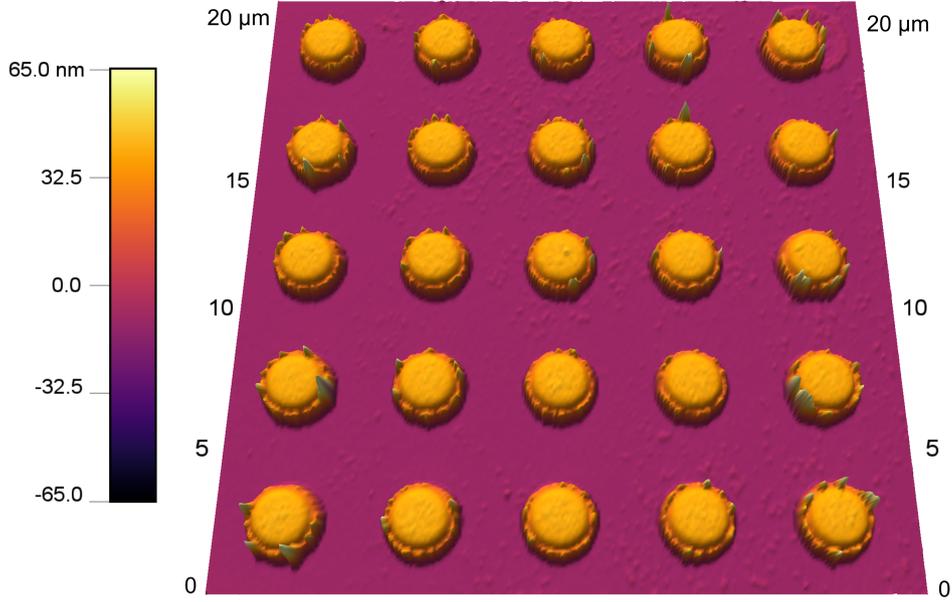

**Supplementary Fig. 1. 3D topography image of 2 μm dots of Ti (10 nm)/Pt (20 nm)/FeCoN (20 nm)**. The mentioned dot stack was grown on top of Si/SiO$_2$ substrate before selective contact deposition.

# II. Intra-fractional Hamming distance considering different sources of randomness

**Supplementary Fig. 2** shows calculated intra-fractional Hamming distance (FHD$_{intra}$) for a 24-bit array (Sample 1, Circuit B) treated at $V_G = -10$ V during $t_1 = 60$ min considering four possible states: right- and left-oriented single-domain (SD) states, and clockwise (CW) and counterclockwise (CCW) vortex states. Obtained FHD$_{intra}$ = 0.505 deviates from the ideal value of 0.75 for four possible states, indicating that the system is not fully stochastic but rather exhibits a probabilistic behaviour. When the analysis is limited to SD orientation and vortex chirality–two binary directionality modes, based on grouping left orientation of the SD state with CCW chirality of the vortex state, and right orientation of the SD state with CW chirality of the vortex state–the resulting average FHD$_{intra}$ = 0.474 (**Fig. 3k** of the main text). This value closely aligns with the theoretical value of 0.5 expected for a random binary sequence, indicating that degaussed states are statistically uncorrelated. Actuating the complementary 18 dots (Sample 1, Circuit A) under the same conditions, and considering only SD orientation and vortex chirality, again confirms fully random behaviour. The resulting FHD$_{intra}$ of 0.493 confirms absence of correlations in orientation or chirality across degaussing cycles (**Supplementary Fig. 3**).

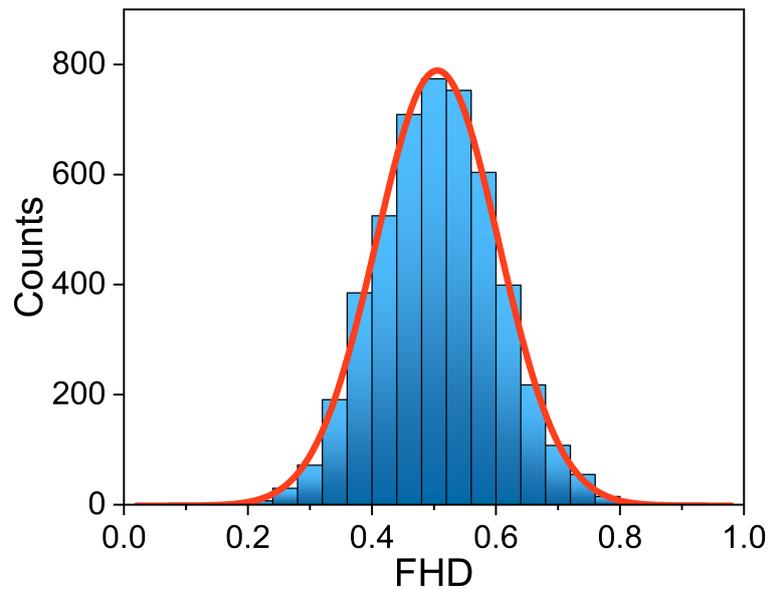

**Supplementary Fig. 2. FHD$_{intra}$ for a 24-bit array (Sample 1, Circuit B) considering four magnetic states: right- and left-oriented SD, and CW and CCW vortex.** The red line indicates a Gaussian fit to the histogram, with an average FHD$_{intra}$ = 0.505, indicating that the probability of obtaining a SD or vortex state after degaussing is unequal.

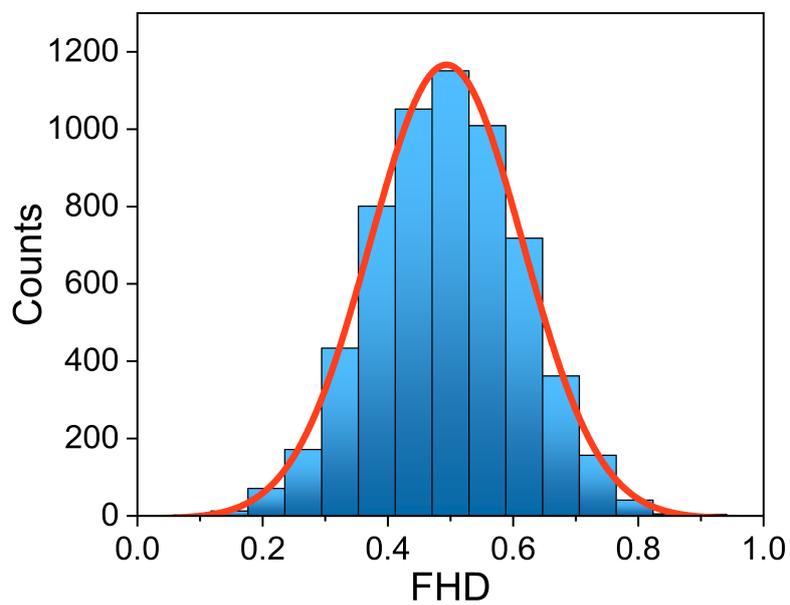

**Supplementary Fig. 3. FHD$_{intra}$ of 18-bit array (Sample 1, Circuit A). considering only the orientation of the SD states and the chirality of the vortex states (*i.e.*, two direction-based subclasses).** The red line is a Gaussian fit to the histogram, yielding an average FHD$_{intra}$ = 0.493, indicating that the degaussed states, when accounting for orientation and chirality, are random and uncorrelated.

## III. Shannon entropy per bit

We calculated the Shannon entropy for each bit in Sample 1 for two circuits: circuit B, composed of $N = 24$ magneto-ionically actuated dots, and circuit A, composed of $N = 18$ dots, as shown in **Supplementary Tables 1,2**, respectively. In both cases, we considered the probabilities $p_i$ of the two sub-classes after degaussing: right (clockwise) or left (counterclockwise) orientation (chirality). The resulting average Shannon entropy per bit was $\overline{H_i} = 0.99$ for circuit A and $\overline{H_i} = 0.97$ for circuit B, indicating a high degree of uncertainty and near-maximum randomness in the distribution of the possible states.

| Dot position | $p_{R,CW}$ (%) | $p_{L,CCW}$ (%) | Shannon entropy/bit |
|---|---|---|---|
| 1 | 61.29% | 38.71% | 0.963 |
| 2 | 32.26% | 67.74% | 0.907 |
| 3 | 53.23% | 46.77% | 0.997 |
| 4 | 35.48% | 64.52% | 0.938 |
| 5 | 46.77% | 53.23% | 0.997 |
| 6 | 64.52% | 35.48% | 0.938 |
| 7 | 38.71% | 61.29% | 0.963 |
| 8 | 35.48% | 64.52% | 0.938 |
| 9 | 50.00% | 50.00% | 1.000 |
| 10 | 45.16% | 54.84% | 0.993 |
| 11 | 33.87% | 66.13% | 0.924 |
| 12 | 43.55% | 56.45% | 0.988 |
| 13 | 30.65% | 69.35% | 0.889 |
| 14 | 50.00% | 50.00% | 1.000 |
| 15 | 46.77% | 53.23% | 0.997 |
| 16 | 45.16% | 54.84% | 0.993 |
| 17 | 35.48% | 64.52% | 0.938 |
| 18 | 61.29% | 38.71% | 0.963 |
| 19 | 61.29% | 38.71% | 0.963 |
| 20 | 41.94% | 58.06% | 0.981 |
| 21 | 38.71% | 61.29% | 0.963 |
| 22 | 48.39% | 51.61% | 0.999 |
| 23 | 53.23% | 46.77% | 0.997 |
| 24 | 48.39% | 51.61% | 0.999 |

**Supplementary Table 1. Shannon entropy per bit for 24 magneto-ionically actuated dots in circuit B (Sample 1).** Shannon entropy was calculated by using the probabilities $p_i$ of two subclasses: right-oriented SD and clockwise CW vortex states ($p_{R,CW}$), and left-oriented SD and CCW vortex states ($p_{L,CCW}$), observed after degaussing. The dots are numbered following the reading direction, from left to right and top to bottom, so that dot number 1 is at the upper-left corner, and dot number 24 is at the bottom-right corner of the corresponding circuit (**Fig. 2f** in the main text).

| Dot position | $p_{R,CW}$ (%) | $p_{L,CCW}$ (%) | Shannon entropy/bit |
|---|---|---|---|
| 1  | 38.71% | 61.29% | 0.963 |
| 2  | 53.23% | 46.77% | 0.997 |
| 3  | 45.16% | 54.84% | 0.993 |
| 4  | 53.23% | 46.77% | 0.997 |
| 5  | 53.23% | 46.77% | 0.997 |
| 6  | 48.39% | 51.61% | 0.999 |
| 7  | 54.84% | 45.16% | 0.993 |
| 8  | 41.94% | 58.06% | 0.981 |
| 9  | 54.84% | 45.16% | 0.993 |
| 10 | 37.10% | 62.90% | 0.951 |
| 11 | 38.71% | 61.29% | 0.963 |
| 12 | 46.77% | 53.23% | 0.997 |
| 13 | 51.61% | 48.39% | 0.999 |
| 14 | 46.77% | 53.23% | 0.997 |
| 15 | 61.29% | 38.71% | 0.963 |
| 16 | 48.39% | 51.61% | 0.999 |
| 17 | 50.00% | 50.00% | 1.000 |
| 18 | 54.84% | 45.16% | 0.993 |

**Supplementary Table 2. Shannon entropy per bit for 18 magneto-ionically actuated dots in circuit A (Sample 1).** Shannon entropy was calculated by using the probabilities $p_i$ of two subclasses: right-oriented SD and CW vortex states ($p_{R,CW}$), and left-oriented SD and CCW vortex states ($p_{L,CCW}$), observed after degaussing. The dots are numbered following the reading direction, from left to right and top to bottom, so that dot number 1 is at the upper-left corner, and dot number 18 is at the bottom-right corner of the corresponding circuit (**Fig. 2e** in the main text).

### IV. Enrolled library of states' probabilities

Tracking the spin configurations of individual dots after successive degaussing cycles enables the creation of an enrolled library of SD and vortex state probabilities. During the enrolment process (*i.e.*, the training phase), 100 degauss/imaging cycles were conducted to evaluate the probabilities of SD and vortex states. These results are presented in **Supplementary Table 3** for $N = 18$ magneto-ionic elements in A-type circuit, subjected to a gate voltage of – 10 V for $t_1$ = 60 min (Sample 1) and $t_2$ = 30 min (Sample 2). A majority voted state is identified as the state that occurs most frequently across a set of repeated degaussed images, *i.e.* the most common state observed for a given bit.

| Circuit A Dot position | Sample 1 pSD (%) | Sample 1 pV (%) | Majority voted state (SD/V) | Sample 2 pSD (%) | Sample 2 pV (%) | Majority voted state (SD/V) |
|---|---|---|---|---|---|---|
| 1 | 0 | 100 | V | 0 | 100 | V |
| 2 | 0 | 100 | V | 0 | 100 | V |
| 3 | 0 | 100 | V | 5.9 | 94.1 | V |
| 4 | 0 | 100 | V | 0 | 100 | V |
| 5 | 0 | 100 | V | 100 | 0 | SD |
| 6 | 0 | 100 | V | 0 | 100 | V |
| 7 | 24.2 | 75.8 | V | 55 | 45 | SD |
| 8 | 0 | 100 | V | 0 | 100 | V |
| 9 | 41.9 | 58.1 | V | 11.8 | 88.2 | V |
| 10 | 11.2 | 88.8 | V | 0 | 100 | V |
| 11 | 0 | 100 | V | 27.4 | 72.6 | V |
| 12 | 0 | 100 | V | 78.4 | 21.6 | SD |
| 13 | 91.1 | 8.9 | SD | 66.6 | 33.4 | SD |
| 14 | 3.2 | 96.8 | V | 0 | 100 | V |
| 15 | 0 | 100 | V | 41.1 | 58.9 | V |
| 16 | 0 | 100 | V | 0 | 100 | V |
| 17 | 15.6 | 84.4 | V | 37.2 | 62.8 | V |
| 18 | 6.5 | 93.5 | V | 85 | 15 | SD |

**Supplementary Table 3. Enrolled library of probabilities for obtaining a SD state (*p*SD) or a vortex state (*p*V) in Sample 1 (circuit A, treated at – 10 V for 60 min) and Sample 2 (circuit A, treated at – 10 V for 30 min).** Deterministic bits are highlighted in cyan and probabilistic in grey colour, respectively. An additional column for each sample labelled as "majority voted state" shows the most probable state of each bit after 100 degaussing/imaging cycles, determined by a majority voting procedure. The dots are numbered following the reading direction, from left to right and top to bottom, so that dot number 1 is at the upper-left corner, and dot number 18 is at the bottom-right corner of circuit A in both samples.

### V. Inference protocol

We implement an inference protocol to determine whether a set of input images involving five *p*-bits, corresponds to Sample 1 or Sample 2. As an example, we acquired successive images from Sample 1 after repeated degaussing and, using a pre-established probability library for both samples (**Supplementary Table 3**), assigned each dot to the more likely sample labelled as "1" or "2". Classification is performed across five *p*-bits using a majority voting scheme. In practice, this protocol can be applied to any unknown sample to verify its authenticity against one or more reference samples.

In a given trial $T_i$, the probability of the sample being identified as Sample 1 is defined as the ratio of dots classified as "1" to the total number of dots (**Supporting Table 4**). While initial iterations may yield probabilities below 50 %, occasionally leading to misclassification (see **Supplementary Table 4** and **Fig. 4k**), the inference process quickly corrects. After four iterations, the cumulative probability–averaging probabilities from all previous trials–exceeds 50 %, reaching 90 % after 27 iterations.

| Dot position | *p*SD (Sample 1) | *p*SD (Sample 2) | $T_1$ | $T_2$ | $T_3$ | $T_4$ | $T_5$ | $T_6$ | $T_7$ | $T_8$ | $T_9$ |
|---|---|---|---|---|---|---|---|---|---|---|---|
| Dot 13 | 91.1 | 66.6 | 2 | 2 | 1 | 1 | 1 | 1 | 1 | 1 | 1 |
| Dot 7 | 24.2 | 55 | 2 | 2 | 2 | 1 | 1 | 2 | 1 | 1 | 1 |
| Dot 9 | 41.9 | 11.8 | 2 | 2 | 1 | 1 | 1 | 1 | 2 | 1 | 1 |
| Dot 18 | 6.5 | 85 | 1 | 1 | 1 | 2 | 1 | 1 | 1 | 1 | 1 |
| Dot 17 | 15.6 | 37.2 | 1 | 1 | 2 | 1 | 1 | 1 | 1 | 1 | 1 |
| Probability per trial (%) | | | 40 | 40 | 60 | 80 | 100 | 80 | 80 | 100 | 100 |
| Cumulative probability (%) | | | 40.0 | | 46.6 | | 64.0 | | 68.5 | | 75.5 |
| Dot position | *p*SD (Sample 1) | *p*SD (Sample 2) | $T_{10}$ | $T_{11}$ | $T_{12}$ | $T_{13}$ | $T_{14}$ | $T_{15}$ | $T_{16}$ | $T_{17}$ | $T_{18}$ |
| Dot 13 | 91.1 | 66.6 | 1 | 1 | 1 | 1 | 1 | 1 | 1 | 1 | 1 |
| Dot 7 | 24.2 | 55 | 1 | 1 | 1 | 1 | 1 | 1 | 1 | 1 | 1 |
| Dot 9 | 41.9 | 11.8 | 2 | 1 | 1 | 1 | 2 | 1 | 1 | 1 | 1 |
| Dot 18 | 6.5 | 85 | 1 | 1 | 1 | 1 | 1 | 1 | 2 | 1 | 1 |
| Dot 17 | 15.6 | 37.2 | 1 | 1 | 2 | 1 | 1 | 1 | 1 | 1 | 1 |
| Probability per trial (%) | | | 80 | 100 | 80 | 100 | 80 | 100 | 80 | 100 | 100 |
| Cumulative probability (%) | | | | 78.2 | | 80.0 | | 81.3 | | 82.4 | |
| Dot position | *p*SD (Sample 1) | *p*SD (Sample 2) | $T_{19}$ | $T_{20}$ | $T_{21}$ | $T_{22}$ | $T_{23}$ | $T_{24}$ | $T_{25}$ | $T_{26}$ | $T_{27}$ |
| Dot 13 | 91.1 | 66.6 | 1 | 1 | 1 | 1 | 1 | 1 | 1 | 1 | 1 |
| Dot 7 | 24.2 | 55 | 1 | 1 | 1 | 1 | 1 | 1 | 1 | 1 | 1 |
| Dot 9 | 41.9 | 11.8 | 1 | 1 | 1 | 1 | 1 | 1 | 1 | 1 | 1 |
| Dot 18 | 6.5 | 85 | 1 | 1 | 1 | 1 | 1 | 1 | 1 | 1 | 1 |
| Dot 17 | 15.6 | 37.2 | 1 | 1 | 1 | 1 | 1 | 1 | 1 | 1 | 1 |
| Probability per trial (%) | | | 100 | 100 | 100 | 100 | 100 | 100 | 100 | 100 | 100 |
| Cumulative probability (%) | | | 84.2 | | 85.7 | | 86.9 | | 88.0 | | 88.9 |

**Supplementary Table 4. Demonstration of the inference protocol for Sample 1 (circuit A) across 27 successive degauss/imaging iterations using five *p*-bits**. Each trial $T_i$ assigns the *p*-bits to Sample 1 or Sample 2 (1 or 2) by comparing them to the pre-

established (*i.e.* enrolled) probability library (**Supplementary Table 3**). The enrolled probabilities of the SD states ($p$SD) in Sample 1 and Sample 2 are also listed in the second and third column of this table. The probability of a vortex state is $p$V (%) = 100 – $p$SD. Initial trials (*e.g.*, $T_1$–$T_3$) may lead to misclassification, as the cumulative probability of inferring a Sample 1, based on majority voting within each trial and averaging over all previous iterations, remains below 50 % during the first three iterations. However, cumulative inference rapidly corrects classification, surpassing 50 % confidence by iteration 4 and reaching 90 % by iteration 27.